\documentclass{article}

\usepackage[nonatbib,preprint]{neurips_2023}

\usepackage[utf8]{inputenc} 
\usepackage[style=numeric-comp]{biblatex}
\usepackage[T1]{fontenc}    
\usepackage{hyperref}       
\usepackage{url}            
\usepackage{booktabs}       
\usepackage{amsfonts}       
\usepackage{nicefrac}       
\usepackage{microtype}      
\usepackage{xcolor}         
\usepackage{graphicx}
\usepackage{amsmath}
\usepackage{bm}

\title{Learning low-dimensional dynamics from whole-brain data improves task capture}

\author{%
  Eloy Geenjaar \\
  TReNDS center \\
  Georgia Institute of Technology \\
  \texttt{egeenjaar@gatech.edu} \\
  \And
  Donghyun Kim \\
  TReNDS center \\
  Georgia Institute of Technology \\
  \And
  Riyasat Ohib \\
  TReNDS center \\
  Georgia Institute of Technology \\
  \And
  Marlena Duda \\
  TReNDS center \\
  Georgia State University \\
  \And
  Amrit Kashyap \\
  Computational Neurology \\
  Charite University Hospital \\
  \And
  Sergey Plis \\
  TReNDS center \\
  Georgia State University \\
  \And
  Vince Calhoun \\
  TReNDS center \\
  Georgia Institute of Technology \\
}

\begin{document}

\maketitle

\begin{abstract}
The neural dynamics underlying brain activity are critical to understanding cognitive processes and mental disorders. However, current voxel-based whole-brain dimensionality reduction techniques fall short of capturing these dynamics, producing latent timeseries that inadequately relate to behavioral tasks. To address this issue, we introduce a novel approach to learning low-dimensional approximations of neural dynamics by using a sequential variational autoencoder (SVAE) that represents the latent dynamical system via a neural ordinary differential equation (NODE). Importantly, our method finds smooth dynamics that can predict cognitive processes with accuracy higher than classical methods. Our method also shows improved spatial localization to task-relevant brain regions and identifies well-known structures such as the motor homunculus from fMRI motor task recordings. We also find that non-linear projections to the latent space enhance performance for specific tasks, offering a promising direction for future research. We evaluate our approach on various task-fMRI datasets, including motor, working memory, and relational processing tasks, and demonstrate that it outperforms widely used dimensionality reduction techniques in how well the latent timeseries relates to behavioral sub-tasks, such as left-hand or right-hand tapping. Additionally, we replace the NODE with a recurrent neural network (RNN) and compare the two approaches to understand the importance of explicitly learning a dynamical system. Lastly, we analyze the robustness of the learned dynamical systems themselves and find that their fixed points are robust across seeds, highlighting our method's potential for the analysis of cognitive processes as dynamical systems.
\end{abstract}

\section{Introduction}
Functional magnetic resonance imaging (fMRI) is a highly informative non-invasive whole-brain modality used to study oxygen-based changes in the brain, which has been essential in understanding cognitive processes~\cite{poldrack2012future}. Alongside EEG, fMRI is one of the only widely used modalities for whole-brain functional imaging, demonstrating its efficacy and importance. Although oxygen-based changes are a proxy signal for neuronal activity, fMRI remains a highly reliable modality to study whole brain function. The analysis of fMRI data is also challenging due to its low signal-to-noise ratio and relatively few training samples compared to its high spatial dimensionality. Researchers have attempted to overcome these challenges using powerful dimensionality reduction techniques. The most prominent dimensionality reduction techniques currently used are averaging/grouping voxels based on a neuroanatomical atlas parcellation~\cite{tzourio2002automated, glasser2016multi, yeo2011organization}, independent component analysis (ICA)~\cite{oja2000independent, mckeown1998independent, calhoun2006unmixing}, and principal component analysis (PCA)~\cite{wold1987principal, smith2014group}. PCA specifically is also often used after neuroanatomical atlas parcellation and is an important preprocessing step for ICA. All three map the functional signal to a temporal trajectory in a low-dimensional subspace. Parcellating neural data, specifically, has drawbacks in representing the neural activity in a region as a single timeseries. Many subregions can act independently during tasks as well as spatially overlap with one another.

On a different scale, work in computational neuroscience aimed at understanding spiking neural data also reduces the dimensionality of their data to understand its underlying computation. Previously, researchers hypothesized that single neurons would be related to specific visual inputs~\cite{quiroga2005invariant} but our contemporary understanding is that cognitive processes are likely more directly related to the lower-dimensional dynamics of groups of neurons~\cite{vyas2020computation}. A similar revelation is currently unfolding at the whole-brain scale. Functional localization is likely not a perfect model for higher cognitive processes from whole-brain data~\cite{pessoa2023entangled}. Due to the unique challenges of fMRI data, we have not yet seen this matched with new low-dimensional dynamical system approaches for fMRI data. Most recent approaches that focus on understanding whole-brain dynamics are \textit{downstream} from the dimensionality reduction step~\cite{misra2021learning, song2022large, iyer2022focal}. Dynamic functional connectivity~\cite{rashid2014dynamic, damaraju2014dynamic, xie2019efficacy} approaches, hidden Markov models~\cite{suk2016state}, or embedding algorithms, such as temporal potential of heat diffusion for affinity-based transition embedding (T-PHATE)~\cite{busch2023multi} typically rely on preprocessing, such as dimensionality reduction techniques or mask voxels using an atlas. Additionally, the lack of a dynamical system makes it difficult to interpret these low-dimensional dynamics and how they relate to behavior. In computational neuroscience, preliminary work shows how powerful dynamical systems theory can be when analyzing learned low-dimensional dynamics. Therefore, with this work, we introduce a new \textit{upstream} method for dimensionality reduction that captures a low-dimensional representation of fMRI dynamics with varying degrees of nonlinearity and is complementary to existing approaches.

Our proposed dynamical system-based approach provides not only a technical advancement but also opens a path for a novel line of analyses and interpretations for future fMRI research. Firstly, we believe that learning low-dimensional dynamical systems can be an especially fruitful avenue for exploring the connection between structural modalities, such as structural MRI (sMRI) and diffusion MRI (dMRI), and the geometry of the manifold learned in the latent space from functional data~\cite{pessoa2019neural}. Since most studies acquire all three modalities from the same subject, functional MRI data is particularly well-suited for this line of research. We hope that this direction can help bridge the gap with more bottom-up approaches~\cite{deco2008dynamic, breakspear2017dynamic, pang2022geometric}. Secondly, dynamical systems can be a powerful new way of studying cognitive processes and mental disorders. We demonstrate that our proposed method both more accurately relates to the task-related cognitive processes and directly models spatially localized task-based variance in the data. Thus, our dynamical approach could act as a complementary approach to the commonly used state-based model of whole-brain activity~\cite{rashid2014dynamic, calhoun2016time}. Lastly, it is relevant to note that our model is not for causal inference; however, finding underlying dynamical models can highlight potential mechanisms for further study through intervention and can inform potentially fruitful future directions for causal inference research in the brain, such as with transcranial magnetic stimulation (TMS)~\cite{bestmann2008mapping}.

Our framework learns a low-dimensional dynamical system directly from whole-brain fMRI data. It is the first framework of its kind for high-dimensional whole-brain data, as far as we are aware. Our goal in this work then is threefold; first, we aim to assess the potential of low-dimensional dynamical systems for fMRI data. Given that for resting fMRI data, there is no direct ground truth of the underlying computations, we apply our method to task fMRI data, where the occurrences of cognitive processes are labeled. We then evaluate how separable and identifiable individual sub-tasks, e.g. left-hand vs left-foot movement are using the latent representation and a linear classifier. This is extremely useful to address in fMRI, as the low signal-to-noise ratio and high spatial dimensionality can obfuscate the identity of a sub-task. However, dimensionality reduction techniques may also project noise into the latent space. This leads to a lower reconstruction error but also information in the latent space that is unrelated to the cognitive process. To ensure that our model captures signal instead of noise, we evaluate how well our model trained on a motor task captures variance for the motor homunculus compared to other common dimensionality techniques. The motor homunculus is a well-studied topographical representation of the motor cortex that maps body parts to specific regions of the cortex and thus acts as a perfect ground truth for the localization of signal vs noise during a motor task. We also exemplify the benefits of explicitly using a dynamical system in the latent space by analyzing the stability of the fixed points of the dynamical system across seeds and tasks. Our second aim is to compare two versions of our method, one with a neural ODE~\cite{chen2018neural} and another with an RNN as the dynamical system in the latent space. This is based on recent work~\cite{sedler2023expressive}, that showed Neural ODEs learn low-dimensional dynamical systems more accurately than RNNs. Lastly, given the low signal-to-noise and high dimensionality of fMRI data, we believe that non-linear projections to the low-dimensional space may ease the learning of the low-dimensional dynamical system. Thus, we compare both linear and non-linear projections to the low-dimensional latent space.

Section~\ref{sec:method} will outline the technical details of our approach, which is visualized in Figure~\ref{fig:figure1}, and we describe our evaluation method and results in Section~\ref{sec:experiments} (Figures~\ref{fig:figure2},\ref{fig:figure3},\ref{fig:figure4}). Section~\ref{sec:background} relates our work to previous research, and we discuss the results, limitations of our current approach, and thoughts on future work in Section~\ref{sec:discussion}.

\section{Background}
\label{sec:background}
Dimensionality reduction is an essential step of fMRI data analysis pipelines, especially when done via sequence-to-sequence models, which map an fMRI trial $\mathbf{X} \in \mathbb{R}^{T \times N}$ with $T$ timepoints and $N$ voxels to a latent timeseries $\mathbf{Z} \in \mathbb{R}^{T \times d}$, where $d \ll N$. In general, we can think of this mapping as a function $h_{\mathbf{\theta}}(\cdot): \mathbb{R}^{T \times N} \to \mathbb{R}^{T \times d}$ parameterized by $\mathbf{\theta}$.

In most existing methods $h_{\bm{\theta}}$ is a linear transformation expressed as a matrix multiplication, such that $h_{\bm{\Theta}}(\mathbf{X}) = \mathbf{X}\mathbf{\Theta} = \mathbf{Z}$.
The most common methods for the analysis of fMRI that operate under this assumption are parcellation with an anatomical atlas~\cite{yeo2011organization, glasser2016multi, tzourio2002automated}, ICA~\cite{oja2000independent, mckeown1998independent, calhoun2006unmixing}, PCA~\cite{viviani2005functional, gotts2020brain}, and dictionary learning~\cite{eavani2012sparse}.
Other dimensionality reduction techniques that instead focus on reducing temporal dimensionality are topographic factor analysis~\cite{manning2014topographic, manning2014hierarchical}, neural topographic factor analysis~\cite{sennesh2020neural}, or point processes~\cite{tagliazucchi2012criticality}.
Given that many anatomical atlas parcellations still result in a high spatial dimensionality $(10-180)$, previous work has also shown the existence of a non-linear low-dimensional manifold from neuro-anatomically parcellated fMRI data \cite{gotts2020brain}. Furthermore, anatomical atlas parcellations have been used in dynamical causal modeling \cite{cao2019functional, zhuang2021multiple}, deep Markov factor analysis \cite{farnoosh2021deep}, and to learn Markovian vs non-Markovian dynamics \cite{zappala2022neural} or the initial conditions for a pre-determined dynamical system \cite{kashyap2021deep}. There have, however, been little to no fundamental improvements in voxel-based dimensionality reduction techniques for fMRI data, especially for techniques that explicitly model its temporal dynamics.

While not yet utilized in whole-brain imaging data, learning low-dimensional dynamics from neural data is rather commonplace for neural spiking data~\cite{sussillo2016lfads, pandarinath2018inferring, keshtkaran2022large}. Recently this has been extended to using neural ordinary differential equations~\cite{chen2018neural} (NODEs) on neural spiking data~\cite{kim2021inferring} or to compare recurrent neural networks (RNNs) and NODEs on simulated spiking data~\cite{sedler2023expressive}. We took the latter work as inspiration to perform a comparison between RNNs and NODEs for fMRI data. Additionally, both our models with an RNN and NODE are novel in the realm of fMRI analysis.

\section{Low-dimensional dynamics from fMRI data}
\label{sec:method}
\begin{figure}[t]
    \centering
    \includegraphics[width=0.8\linewidth]{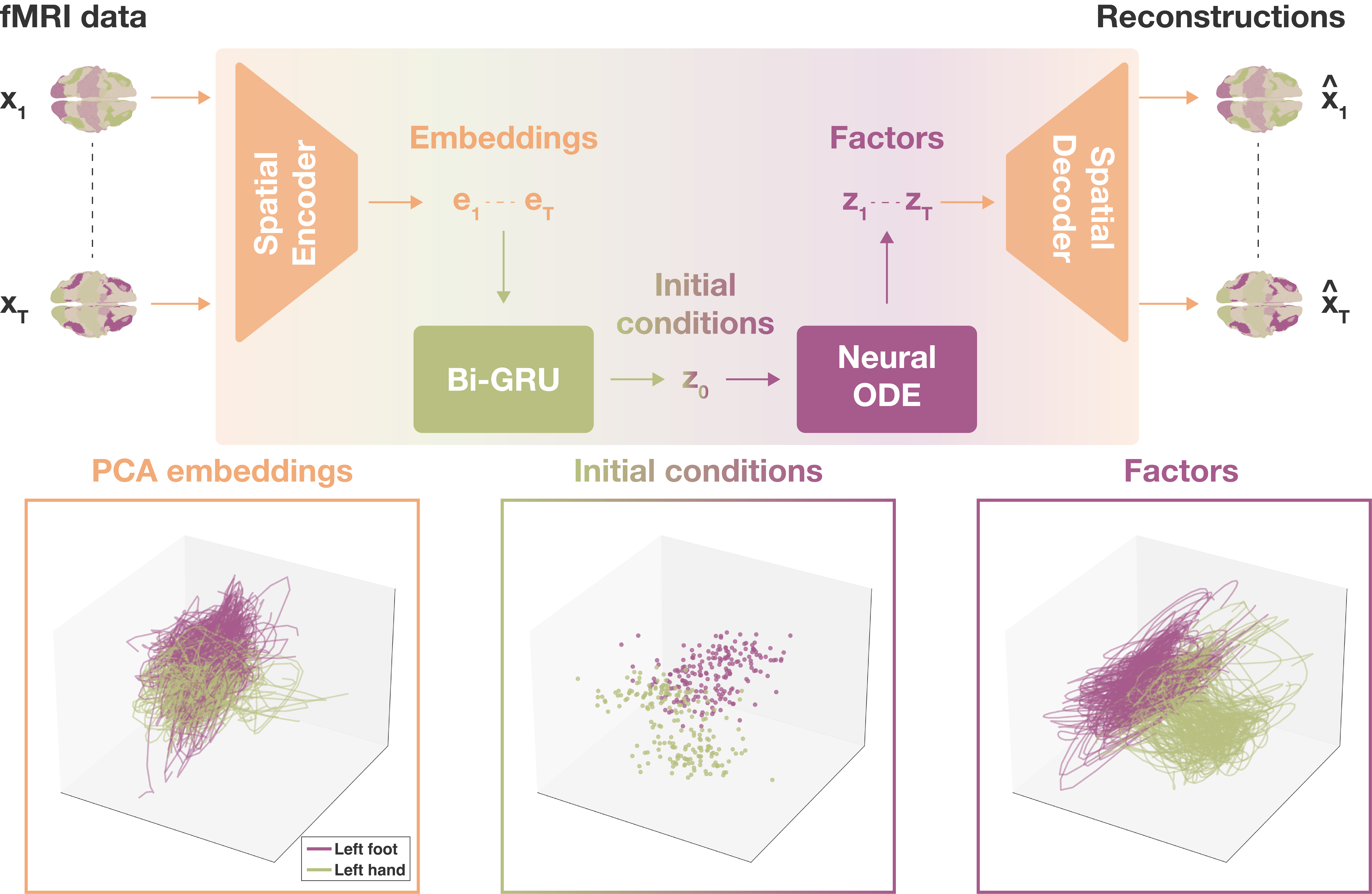}
    \caption{Our model architecture, with example latent timeseries for PCA, the initial conditions from our model, and the factors from our model. The timeseries correspond to a model trained on both left-foot and left-hand tapping task fMRI data.}
    \label{fig:figure1}
\end{figure}

Especially for neuronal data, there is a move towards learning dynamical systems from a low-dimensional manifold of neural activity~\cite{duncker2021dynamics}. Similar evidence of a low-dimensional manifold is emerging for fMRI data~\cite{gotts2020brain}. To learn the dynamics in this low-dimensional latent space, we parameterize a latent autonomous dynamical system of the following form. Autonomous dynamical systems are dynamical systems that do not require any inputs, and given an initial state can completely unroll the temporal dimension of the data.
\begin{align*}
  \dot{\mathbf{z}}_{t} &= \text{F}_{\bm{\theta}}(\mathbf{z}_{t})
\end{align*}
Where $\mathbf{z}_{t}$ is the latent vector at timestep $\text{t}$ and
$\text{F}_{\bm{\theta}}$ parameterizes the time evolution of the
latent vectors. We discretize the latent vectors based on what time
each fMRI volume is acquired. For $\text{F}_{\bm{\theta}}$, we use a
gated recurrent unit (GRU)~\cite{cho2014learning} and a neural
ordinary differential equation (NODE)~\cite{chen2018neural}. Specifically, we can rewrite a discretized
version of the equation as a mapping from $\mathbf{z}_{t}$ to
$\mathbf{z}_{t+1}$ as follows.
\begin{align*}
\mathbf{z}_{t+1} &= \mathbf{z}_{t} + \dot{\mathbf{z}_{t}} \\
                 &= \mathbf{z}_{t} + \text{F}_{\bm{\theta}}(\mathbf{z}_{t})
\end{align*}
For the GRU, we can write this based on the hidden state of the GRU The GRU's hidden state dimensionality, however, needs to be bigger than the latent dimension to effectively learn the dynamics~\cite{sedler2023expressive}. Low-dimensional dynamics often emerge from high-dimensional RNNs~\cite{sussillo2013opening}, so we use a linear mapping to obtain the latent vector at each timestep from the hidden
state, as follows.
\begin{align*}
\mathbf{h}_{t+1} &= \text{GRUCell}(\mathbf{h}_{t}) \\
\mathbf{z}_{t+1} &= \mathbf{h}_{t+1} \mathbf{W}_{z} + \mathbf{b}_{z} \\
&= \mathbf{z}_{t} + \text{F}_{\bm{\theta}}(\mathbf{z}_{t}) \\
\text{F}_{\theta}(\mathbf{h}_{t}) &= \text{GRUCell}(\mathbf{h}_{t})
                                    \mathbf{W}_{z} + \mathbf{b}_{z} -
                                    \mathbf{z}_{t}
\end{align*}
For the NODE, we parameterize $F_{\theta}(\cdot)$ as a multi-layer
perceptron (MLP), and obtain the following equation.
\begin{align}
\label{eq:NODE}
    \mathbf{z}_{t+1} &= \mathbf{z}_{t} + \int_{t}^{t+1} \text{MLP}(\mathbf{z}_{t}) \text{dt}
\end{align}
We point out two important differences between the GRU and NODE. First, the GRU requires a larger hidden dimensionality for its state updates and does not directly update the latent vector. Second, because the NODE parameterizes the derivative instead of the next time step, it uses numerical solvers to calculate the integral in Equation~\eqref{eq:NODE}, leading to smoother dynamics that can interpolate time points with a higher sampling rate than the data.

To model the autonomous dynamics of each fMRI task, the only vector we require is the initial state $\mathbf{z}_{0}$. These initial conditions, together with $\text{F}_{\bm{\theta}}$ completely determine the low-dimensional dynamics. This also allows us to generate more timeseries data in the latent space, purely based on interpolating between different initial conditions. To learn this initial condition, we follow~\cite{pandarinath2018inferring, sedler2023expressive} and use a bi-directional GRU. Due to the high dimensionality of the data (~$90$k spatial dimensions), we embed the original data ($\mathbf{x}_{t} \in \mathbb{R}^{N}$) into an embedding vector ($\mathbf{e}_{t} \in \mathbb{R}^{d}$) with the same dimensionality as the final latent vectors. This embedding is created by the spatial encoder, which in its simplest form is just a single linear layer. To map the latent vectors back to the original space, we use a spatial decoder, see Figure~\ref{fig:figure1}. Both to regularize the network~\cite{kingma2013auto, chen2016variational, su2018variational} and to impose structure onto the initial conditions, we sample it from a variational distribution. So instead of learning $\mathbf{z}_{0}$ directly from the final hidden state of the bi-directional GRU, we learn a mean $\mathbf{\mu}_{0}$ and $\mathbf{\sigma}_{0}$ that together parameterize a normal distribution. We then sample the initial condition from this normal distribution ($\mathbf{z}_{0} \, \sim \, \mathcal{N}(\bm{\mu}_{0}, \bm{\sigma}_{0})$). Similar to a normal variational autoencoder (VAE), we train our sequential variational autoencoder with a reconstruction loss and a KL-divergence loss. The KL-divergence loss is between the distribution of the initial condition distribution and a zero-mean unit-norm normal distribution ($\text{KL}\left(\mathcal{N}(\bm{\mu}_{0}, \bm{\sigma}_{0}) \, || \, \mathcal{N}(\mathbf{0}, \mathbf{1}) \right)$).

As mentioned previously, the simplest form of the spatial encoder and decoder is just a linear layer, that maps from the input data $\mathbf{x}_{t} \in \mathbb{R}^{N}$ to an embedding $\mathbf{e}_{t} \in \mathbb{R}^{d}$ or from a latent vector $\mathbf{z}_{t}$ to a reconstruction $\hat{\mathbf{x}}_{t}$. Given the noise and high dimensionality of the signal, we vary the complexity of both the spatial encoder and the decoder. Our intuition is that adding more non-linearities can help improve the performance of our model by allowing the model to learn a non-linear manifold. Increasing the complexity of the decoder can, however, easily lead to overfitting and a reduction of the variance modeled by the dynamical system~\cite{chen2016variational}. To try and avoid this, we add a skip connection to the non-linearity. The spatial encoders and decoders in our experiments are symmetric, and non-linear versions of our models use a single GELU non-linearity~\cite{hendrycks2016gaussian}.

\section{Experiments}
\label{sec:experiments}
\subsection{Training}
\vspace{-6pt}
\paragraph{Dataset} To ensure we use minimally pre-processed whole-brain data, we use task fMRI data from the Human Connectome Project~\cite{van2013wu}. The dataset consists of $1080$, $1083$, and $1040$ subjects for the motor, relational, and working memory tasks, respectively. We use surface data, with $N=91282$ voxels for each task, except for the task where we only use visual data, which has $N=8788$ voxels. For the motor task, subjects are tasked with tapping either their left or right fingers, squeezing their left or right toes, or moving their tongue. These motor blocks are preceded by a visual cue that tells the subject what body part they should move. Each motor block is $12$ seconds, and each visual cue is $3$ seconds. For the working memory task, the subjects receive a $2.5$-second visual cue informing them of the task type, and for the $0$-Back memory condition, this cue also shows the target. Then, subjects are tasked with either remembering the target ($0$-Back) or whether the picture they see is the same picture from the $2$-Back condition (i.e., $2$ images prior). These two sub-tasks are done in independent blocks of $25$ seconds, each block has $10$ $2.5$-second sub-blocks. In total, there are $8$ larger blocks, four for $0$-Back and four for $2$-Back. These blocks can be subdivided by target type: a tool, body, face, or place. Lastly, for the relational task, the subjects see two pairs of objects, one at the top and one at the bottom of the screen. They first need to decide how the top pair differs (either in shape or texture). Then, subjects should determine whether the bottom pair also differs similarly. This block is called the relational trial. For the control block, the subjects are shown "shape" or "texture" on the screen, and only one object is at the bottom of the screen. The subjects should determine whether the bottom object matches any of the top two objects in terms of the word. Each block lasts $18$ seconds, with $4$ $3.5$-second sub-blocks, with $500$ms between them, for the relational blocks, and $5$ $2.8$-second sub-blocks, with $400$ms between them, for the control blocks. We chose these tasks because the motor task has well-defined ground truth spatial localization, and the other two are complex cognitive processes, thus making for a harder classification task. To train the model, we split the dataset into a training ($70$\%), validation ($10$\%), and test set ($20$\%) that was held out until the final evaluation. To train the models, we separate the timeseries into non-overlapping windows ($23$, $41$, and $27$ timesteps for the motor, working memory, and relational task, respectively) that are as long as the minimum time between two sub-tasks. We perform the windowing because our model assumption is that it is an autonomous system without any inputs, within the windows there are no inputs, but the visual cue itself during the full timeseries is an input to the system. We train separate models for each task and evaluate the models based on the sub-task label belonging to each window. The model is not privy to these labels during training.

\paragraph{Model} The models are trained for $500$ epochs, with $4$ different seeds, a learning rate of $0.001$ for the RNN and NODE, and if the performance had not improved for $50$ epochs, training was stopped. Each model was trained with $2$, $4$, $8$, $16$, $32$, and $64$ latent dimensions (the size of $\mathbf{e}_{t}$ and $\mathbf{z}_{t}$). Both models were also trained with and without a non-linear spatial encoder and decoder, and the hidden size for the encoder and decoder was $128$, as was the hidden size of the bi-directional GRU, single-layer MLP with Tanh activation for the NODE, and RNN in the case of the RNN model. The RNN used in the spatial decoder for the RNN model is a GRU. All models were trained on an internal cluster with NVIDIA A40 GPUs. Training times ranged from about $45$ to $180$ minutes.

\paragraph{Baselines} We compare our model against the most common dimensionality reduction technique, PCA. We performed PCA on the training set and validation set and then used the PCA mapping to find the latent timeseries on the test set. Furthermore, to determine whether the NODE or RNN are necessary, we also evaluated a linear and non-linear version of a VAE. We use the same training parameters as for the NODE and RNN-based models, except we use a learning rate of $5E-5$. Furthermore, the VAE is trained by treating each timestep independently, i.e. without any explicit temporal information.

\subsection{Sub-task classification}
\label{sec:sub-task}
\begin{figure}[ht]
    \centering
    \includegraphics[width=\linewidth]{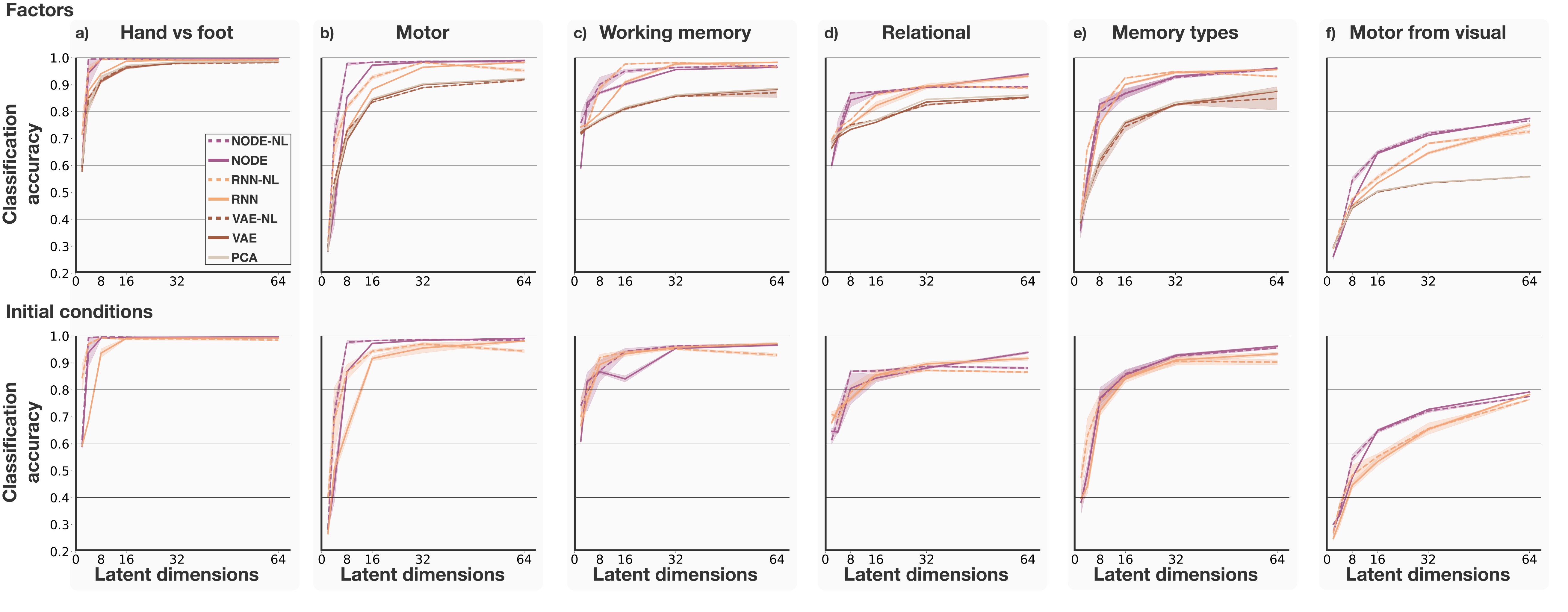}
    \caption{Sub-task classification accuracy from easy (left) to hard (right). Both our models, and even their initial conditions, outperform more common dimensionality reduction techniques, especially with increasing difficulty.}
    \label{fig:figure2}
\end{figure}

\vspace{-6pt}
\paragraph{Experiment \& Evaluation}
The first experiment evaluates how well the latent timeseries relates to the cognitive process evoked by the tasks being performed in the scanner. To do this, we train a logistic regression classifier on each timestep independently and calculate the average classification accuracy across time. Given that our model produces initial conditions, we also assess whether the initial condition alone is a good predictor of the cognitive process. We would expect that different cognitive processes have different dynamics, and the dynamics produced by the NODE or RNN are captured in the initial condition vector. Intuitively, although initial conditions have a very mechanistic use, in our model they also summarize the dynamics into one single vector. This is because the bi-directional GRU in our model sees the full embedded time window before predicting the initial condition. We define different cognitive processes as different sub-tasks performed and we perform six different classification evaluations across the three datasets. We split the motor task into three classification results: first, we simplify the problem to include only left-hand or left-foot tapping sub-task blocks and test the model discriminability between the two (Figure~\ref{fig:figure2}a). We also compare the classification of all five sub-tasks: left hand, left foot, right hand, right foot, and tongue tapping (Figure~\ref{fig:figure2}b). Lastly, we use voxels only from the visual area and predict which of the five motor sub-tasks is being performed (Figure~\ref{fig:figure2}f). Since subjects receive a visual cue in the scanner indicating the upcoming sub-task, we hypothesized dynamics stemming from the visual region alone may sufficiently encode the task at hand. There are two main reasons why this is potentially possible; first, because the visual cue induces different dynamics for each sub-task and/or second because we pick up dynamics from motor-visual feedback connections. The working memory task is also split up into two different classification evaluations. First, we classify if a timeseries is a $0$-Back or $2$-Back block (Figure~\ref{fig:figure2}c), and second we classify what visual element subjects need to remember; places, bodies, faces, or tools (Figure~\ref{fig:figure2}e). Lastly, for the relational processing task, we only evaluate whether a timeseries is a relational or control block (Figure~\ref{fig:figure2}d).

\paragraph{Findings}
Based on Figure~\ref{fig:figure2}, it is clear that our method outperforms PCA, the VAE, and the non-linear VAE across the classification tasks. Furthermore, the NODE-based dynamical system outperforms the RNN-based dynamical system on the motor, 'hand vs foot', and 'motor from visual' tasks, and at low latent dimensionalities for the relational task. Additionally, the non-linear projection generally improves performance for the RNN-based model more clearly than the NODE-based model, although both benefit from it. Due to some of the potential issues we mentioned earlier, it is harder to train the NODE-based model with these non-linear projections, but especially for the motor, working memory, and 'motor from visual' evaluations, the non-linear projection improve the classification accuracy. Lastly, the performance of the initial conditions is often extremely close, or even higher than the factor's average performance across time. 

\begin{figure}[ht]
    \centering
    \includegraphics[width=\linewidth]{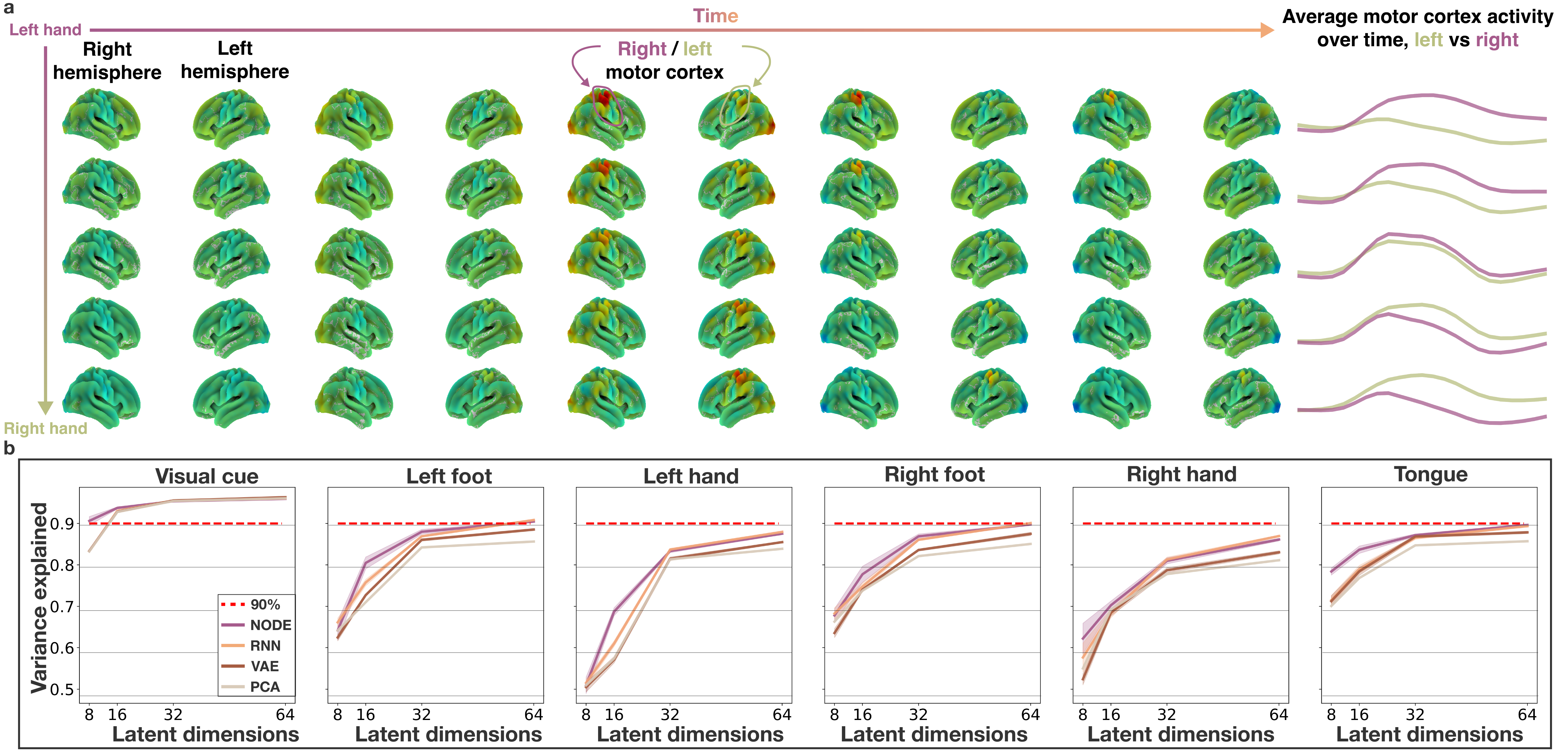}
    \caption{An interpolation between initial conditions in subfigure a, exhibiting interesting dynamic interpolations (right). Variance explained with respect to the motor group maps in subfigure b.}
    \label{fig:figure3}
\end{figure}

\subsection{Spatial specificity}
\label{sec:spatial-specificity}
\vspace{-6pt}
\paragraph{Experiment \& Evaluation} The previous results evaluate whether the dynamics in the latent space are related to each of the sub-tasks across a variety of classification assessments. This raises the question of whether the transformation from the voxel space to the latent space itself is more task-specific as well. To understand if this is the case, we compare the version of our model with a linear projection to the baseline models. For each of the models, we end up with a trained matrix $\mathbf{W}_{\text{decoder}} \in \mathbb{R}^{d \times N}$, indicating the linear mapping between each latent dimension (out of $d$ dimensions) and the voxels in the brain. To understand how well this transformation captures specific areas of the brain, we linearly regress each voxel to a brain map representing the motor homunculus. In this linear regression, each latent dimension is a 'feature', since for PCA and our model, the latent dimensions are not necessarily independent. For both the working memory and relational processing tasks, although there are regions that have been indicated for the task. It is more likely that there are distributed processes throughout the whole brain that contribute to the dynamics of these higher-level cognitive processes. For the motor task, however, the functional specialization in the motor region for specific parts of the body is well known. This topographic specialization is called the motor homunculus. The dataset provides a version of the motor homunculus in the form of task-fMRI Cohen's d group average maps. For each sub-task, we threshold all values in the map below $0.2$ to retain robust spatial locations. We then calculate what quantile of voxels are masked out, and mask out the same number of voxels for the weight matrices in our models ($\mathbf{W}_{\text{decoder}}$). Finally, we fit a linear regression model between each sub-task's group average map and each model's weight matrix. We then calculate the variance explained in the sub-task's group average map and use it as a measure of correct spatial specificity. A high variance explained indicates that the transformation from the latent space to the group average map does focus on the correct robust spatial locations. We only show results for $8$ latent dimensions because the variance explained for $2$ and $4$ latent dimensions is extremely low. We also include an example of interpolation between the mean initial conditions for two sub-tasks in Figure~\ref{fig:figure3}a, to demonstrate the high reconstruction quality of our model and interpolation as a way to perform interpretability analyses on the model.

\paragraph{Findings}
In Figure~\ref{fig:figure3} we show an interpolation between reconstructions of left-hand to right-hand dynamics. The only input we vary is the initial condition we provide the NODE. Since the initial conditions are trained with a variational loss, the manifold they exist on is fairly smooth, enabling realistic interpolations. Note that the left part of the body is represented in the right hemisphere, and vice versa. The spatial specificity results in Figure~\ref{fig:figure3}b show that the NODE is  better than the RNN at low dimensionalities, but outperforms PCA and the VAE. 

\begin{figure}[ht]
    \centering
    \includegraphics[width=0.75\linewidth]{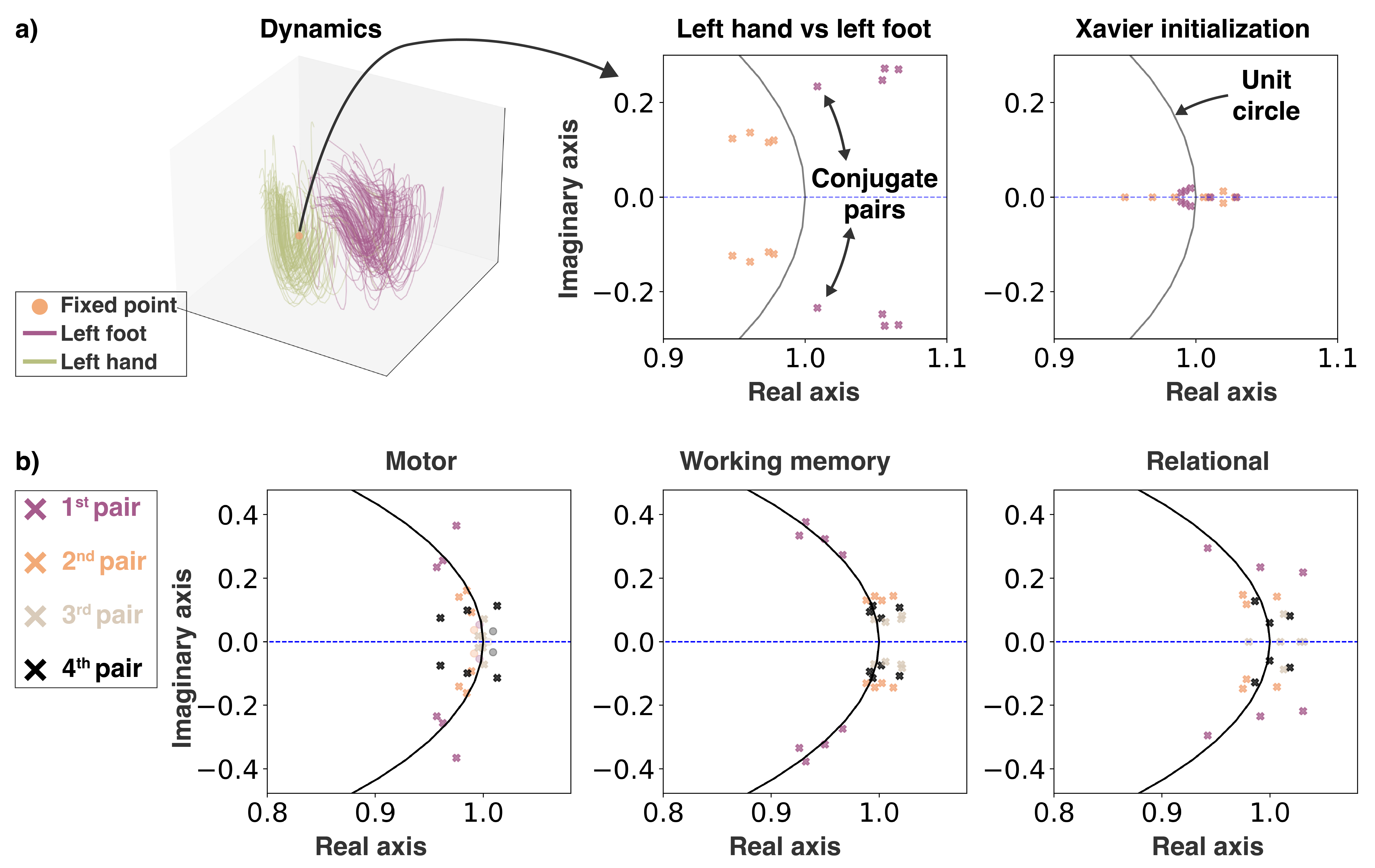}
    \caption{Subfigure a visualizes how we go from a fixed point in the latent space (left) to eigenvalues (middle) by linearizing around the fixed point and what the eigenvalues look like at initialization (right). We repeat each run for four different seeds. Subfigure b shows the eigenvalues for the fixed point we find for motor, working memory, and relational. The goal of subfigure b is to assess the robustness of the fixed points.}
    \label{fig:figure4}
\end{figure}

\subsection{Fixed point robustness analysis}
\label{sec:fixed-point}
\vspace{-6pt}
\paragraph{Experiment \& Evaluation} An especially unique aspect of our method is the ability to analyze the behavior of the learned dynamical system. The most common approach to understanding a dynamical system is to look at points where the derivative is zero, the so-called fixed points. These fixed points are learned in our model during training (See Figure~\ref{fig:figure4}a), and by linearizing around them to calculate the Jacobian at the fixed point, we can analyze the behavior of the system through the eigenvalues of the Jacobian (See Figure~\ref{fig:figure4}). For example, in the work that we use to model our fixed point analysis~\cite{sussillo2013opening}, the authors reconstruct sine waves with varying frequencies using an RNN. They find that each frequency corresponds to a different fixed point and that the frequency of the sine wave it is reconstructing corresponds exactly to the imaginary part of its eigenvalues, i.e., how fast trajectories rotate around the fixed point. In our case, since neural data is much noisier, we only focus on whether it is even possible to find robust fixed points from fMRI data, both to highlight the importance of training our model with multiple seeds, but also because making direct inferences about these eigenvalues requires a more in-depth study. To find the fixed points, we train our model on longer time windows, as it allows the model to converge to a more stable solution. Specifically, in each task there are multiple fixation blocks where subjects do not do any task and focus on a dot, thus we define our longer windows as the sub-task block plus the full subsequent fixation block to train the model. This allows our model to see the full hemodynamic response. Our methodology uses previous work~\cite{sussillo2013opening} that led to a software package~\cite{golub2018fixedpointfinder}, which was converted to PyTorch for recent work~\cite{sedler2023expressive}. In Figure~\ref{fig:figure4}a, we first show the reader an illustrative example of a fixed point for the left hand vs left foot task from our $4$-dimensional models. We show the eigenvalues of the fixed point and compare them to the eigenvalues of the fixed point at initialization. Then, in Figure~\ref{fig:figure4}b, we perform the robustness analysis, but for the longer sequences, and a larger model size ($8$ dimensions). All analyses use the NODE-based system because it produces more accurate eigenvalues on simulation data~\cite{sedler2023expressive}. A larger latent space makes it harder to find the fixed point because the search space is larger, and the tasks in Figure~\ref{fig:figure4}b are more complex. Thus, we stress how non-trivial it is to obtain the same eigenvalues across models trained with a different seed.

\paragraph{Findings} For each seed, the model dimensions are sorted by the magnitude of the imaginary part of the eigenvalue (higher magnitude = higher rank), depicted by the color of each point in Figure~\ref{fig:figure4}. All of the eigenvalues come in conjugate pairs, and each pair appears $4$ times, one for each seed. Thus, clustering of the same color in the figure corresponds to more robust fixed points. Furthermore, if the eigenvalues lie within the unit circle, they correspond to stable oscillations, and eigenvalues outside correspond to unstable oscillations, where the imaginary part indicates the frequency of that oscillation. For the motor task, one model did not converge well, its eigenvalues are shown as circles in~\ref{fig:figure4}b, and the last fixed point did not converge for the relational task, so we did not obtain any eigenvalues. The eigenvalues of the working memory task seem to be tightly clustered and quite robust. For the relational task, the first conjugate pair always seem to capture roughly the same frequency (lie on the same y-axis), and the fourth pair for the motor task exhibits similar behavior. Although for the simpler example (Figure~\ref{fig:figure4}a), eigenvalues do not cross the unit circle across seeds, this does seem to happen for some of the eigenvalues in the motor and relational tasks, especially. Overall, the eigenvalues are surprisingly robust, and we believe this approach is suitable for a more in-depth analysis of our model.

\section{Discussion}
\label{sec:discussion}

Understanding the neural dynamics that underlie brain activity is crucial to unravel cognitive processes. In this study, we propose a novel approach for whole-brain fMRI data, namely to use a sequential variational autoencoder (SVAE). We compare NODE and RNNs as the dynamical system in our model. To validate the efficacy of our approach, we conducted extensive evaluations on various task-fMRI datasets, including motor, working memory, and relational processing tasks. Our method consistently outperformed widely used dimensionality reduction techniques in terms of classification accuracy of the sub-task. We believe our model outperforms the other methods because it directly learns smooth dynamics for the fMRI data, and is thus constrained insofar that it is less likely to learn noise. This hypothesis is further strengthened by our second set of results, where we show that our model learns latent factors that correspond more directly to the motor homunculus than the other methods. The ability to localize brain regions associated with specific cognitive processes further enhances the interpretability and utility of our method in cognitive and mental health research. Furthermore, we compared both NODE and RNN-based models. Especially for the motor and 'motor from visual' tasks, the NODE outperformed the RNN for lower latent dimensionalities. For most other tasks, the performance was roughly the same. Interestingly, in our comparison of non-linear and linear spatial encoders and decoders, we found that the non-linear projection often improved classification accuracy over the linear projection, especially for lower latent dimensionalities. This result occurred for both the RNN and NODE-based models, although more clearly for the RNN, but did not occur for the VAE baseline. Finally, we visualized how robust eigenvalues for fixed points are in these models, even when trained on noisy neural data. We find that the eigenvalues of the fixed points are surprisingly stable, which opens up interesting future avenues into dynamical stability from fMRI data. Specifically, we believe the higher imaginary parts for the working memory task correspond to the task design. Namely, each sub-block consists of $10$ quick $2.5$-second questions, which would explain why the frequency of the fixed point is higher than for the motor and relational tasks.

In conclusion, our study presents a novel approach that integrates neural dynamics into dimensionality reduction techniques for analyzing whole-brain fMRI data. This novel approach better captures cognitive processes.
\vspace{-6pt}
\paragraph{Limitations}
An important limitation of the current work is that it is only applied to task fMRI data. We did this explicitly to assure a ground truth, but acknowledge that a significant portion of the field works with resting-state fMRI. Furthermore, although we have tried to address some of the potential issues with a non-linear decoder, we believe there is more work that can be done to ensure it does not reduce variance in the latent timeseries. Specifically, we see that using a non-linear projection improves performance for the RNN. This is not entirely mirrored for the NODE-based model and we believe this is due to potential issues, discussed in Section~\ref{sec:method}. Lastly, using more than four seeds would have strengthened the work, but this was limited due to time and computing constraints.

\newpage

\section{Acknowledgments}
Acknowledgments will be made available after the blind review process.

\printbibliography
\newpage

\end{document}